\documentclass[iop, apjl, twocolappendix]{emulateapj}
\usepackage{graphicx}% Include figure files
\usepackage{bm}% bold math
\usepackage{amsmath,amssymb}
\usepackage{natbib}
\usepackage{float}
\usepackage{subfigure}
\usepackage{color}
\usepackage[caption=false]{subfig}
\usepackage[normalem]{ulem}
\newcommand{\Z}{\mathcal{Z}}
\newcommand{\M}{\mathcal{M}}
\renewcommand{\L}{\mathcal{L}}

\begin{document}
\title{X-ray afterglows of Short gamma-ray bursts: Magnetar or Fireball?}
\author{Nikhil Sarin\altaffilmark{1}, Paul D. Lasky\altaffilmark{2}, and Greg Ashton\altaffilmark{3}}
\shortauthors{Sarin et al.}
\affil{School of Physics and Astronomy, Monash University, Vic 3800, Australia\\
OzGrav: The ARC Centre of Excellence for Gravitational-wave Discovery, Clayton, Victoria 3800, Australia}
\altaffiltext{1}{nikhil.sarin@monash.edu}
\altaffiltext{2}{paul.lasky@monash.edu}
\altaffiltext{3}{greg.ashton@monash.edu}
\begin{abstract}
    The origin of the X-ray afterglows of gamma-ray bursts has regularly been debated. We fit both the fireball-shock and millisecond-magnetar models of gamma-ray bursts to the X-ray data of GRB 130603B and 140903A.  We use Bayesian model selection to answer the question of which model best explains the data. This is dependent on the maximum allowed non-rotating neutron star mass $M_{\textrm{TOV}}$, which depends solely on the unknown nuclear equation of state. We show that the data for GRB140903A favours the millisecond-magnetar model for all possible equations of state, while the data for GRB130603B favours the millisecond-magnetar model if $M_{\textrm{TOV}} \gtrsim 2.3 M_{\odot}$. If $M_{\textrm{TOV}} \lesssim 2.3 M_{\odot}$, the data for GRB130603B supports the fireball-shock model. We discuss implications of this result in regards to the nuclear equation of state and the prospect of gravitational-wave emission from newly-born millisecond magnetars. 
    \end{abstract}

\section{\label{sec:intro}Introduction}
The coincident observation of short gamma-ray burst GRB170817A \citep{GW170817A_GRB} and gravitational waves from a binary neutron star merger \citep{GW170817} confirmed the association between the compact object progenitor model and short gamma-ray bursts. Short gamma-ray bursts are often followed by an extended emission in lower energy electromagnetic bands referred to as an afterglow. The origin of the afterglow, particularly the X-ray afterglow, is a source of debate. 
Some models attribute it to an expanding fireball that emits X-ray photons through synchrotron emission once the jet hits the surrounding interstellar medium \citep{Meszaros1993, Piran1998, Meszaros1999, Zhang2006}, while others attribute it to a combination of an expanding fireball and a millisecond spin-period magnetar central engine \citep{Dai1998, Zhang2001,Fan2006, Rowlinson2010, Rowlinson2013}. 
In this work, we consider the question of which model best explains the data for two short gamma-ray bursts: GRB130603B and GRB140903A. The component of the afterglow corresponding to the fireball-shock model is always believed to be present and produces an effect on the afterglow in several electromagnetic bands while the millisecond-magnetar model provides an additional dominant component to the X-ray afterglow. 

GRB130603B is believed to be the first credible detection of a kilonova associated with a short gamma-ray burst \citep{Tanvir2013, Berger2013} observed first by the Neil Gehrels \textit{Swift} observatory and then with \textit{XMM} up to 6.5 days after the initial burst \citep{Fong2014}. The millisecond magnetar model has been used extensively to explain the X-ray plateau observed in this gamma-ray burst \citep{Fan2013, DeUgarte2014, Fong2014, Lu2015, LaskyLeris2017}. 

GRB140903A was observed by \textit{Swift} on the 3rd of September 2014 with follow-up observations with \textit{Chandra} $\approx 3$ and $\approx 15$ days after the initial burst \citep{Troja2016}. This gamma-ray burst is especially intriguing from our perspective as both the fireball-shock and millisecond magnetar models have been successfully fit to the observations, with no conclusion available for which model best describes the data. \cite{Troja2016} fit the fireball model to this gamma-ray burst using the X-ray and other wavelength observations, inferring a narrow jet opening angle of $\theta \approx 5^{\circ}$ among other physical quantities such as the jet-break times, suggesting that the observations could be a product of the jet geometry and dynamics within the fireball model. However, both \cite{Zhang2017} and \cite{LaskyLeris2017} successfully fit the millisecond magnetar model to the same X-ray observations.

GRB140903A highlights the need for systematic model selection between the fireball and millisecond magnetar model.  In this paper, we use Bayesian inference and model selection to show which of the two models best explain the data for the two aforementioned GRBs.  In Secs. \ref{sec:FB} and \ref{sec:magnetar} we introduce the fireball and magnetar models, respectively.  In Sec. \ref{sec:mds} we compare our results for GRB130603B and GRB140903A and select between the fireball and millisecond-magnetar models with an {\it uninformed} and an {\it informed} prior odds. The latter being based on the probability that a long-lived millisecond magnetar is born in the gamma-ray burst. 
We discuss the implications of our result on the prospect of gravitational-wave detection and neutron star equation of state in Sec. \ref{sec:conclusion}.
%%%%%%%%%%%%%%%%%%%%%%%%%%%%%%%%%%%%%%%%%%%%%%%%%
\section{Fireball model}\label{sec:FB}
The fireball-shock model has been successful in interpreting a large fraction of gamma-ray bursts. In this model, the ejecta is composed of several shells of matter with a distribution of Lorentz factors. 
The relativistic fireball sweeps through the ambient interstellar medium which decelerates the fireball producing a pair of shocks; a long-lived forward shock and a short-lived reverse shock. The former shock produces the broadband afterglow \citep{Piran1998, Meszaros1993, Meszaros1999, Meszaros2001}. 
\cite{Sari1999} determined the light-curve signature for synchrotron emission from a power-law distribution of accelerated electrons produced by the long-lived shock. The shape and evolution of this light-curve strongly depends on the frequency of the synchrotron emission as well as the geometry of the fireball itself. 
At high frequencies above the self-absorption frequency of lower energy electrons, the flux density in a given frequency band $\nu$, can be parameterized by the electron power-law distribution index $p$ as $F_{\nu} \propto t^{-3(p-1)/4}$ for a spherical fireball. 
However, generally, a parameter-free description of the form 
\begin{equation}\label{Eq.Fb}
F_{\nu} \propto t^{\alpha}\nu^{\beta},
\end{equation}
is used \citep[e.g.,][]{Sari1999, Zhang2006}. Here, $t$ is the time since burst, $\alpha$ is the temporal index, and $\beta$ is the spectral index. The fireball model is characterized by a series of power laws of this form \citep[e.g.,][]{Zhang2006}. 

In reality, there is a strong physical relationship between the temporal and spectral indices based on the properties of the surrounding environment, such as interstellar density. 
However, in this paper we only model the X-ray component in a single frequency band between 0.2 and 10 keV corresponding to the energy range of \textit{Swift}, in which case Eq. (1) can be re-expressed in terms of the luminosity as
\begin{equation}
L(t) = At^{\alpha}.
\end{equation}
Here, $L$ is the luminosity, and $A$ is an amplitude that incorporates the frequency-dependent scaling term from Eq. (\ref{Eq.Fb}). Since we are only looking at the temporal evolution a change in temporal index alone may not be indicative of a jet-break. We focus solely on the X-ray observations in this paper to allow direct comparison with the millisecond-magnetar model. 
We elaborate on other frequency bands later but note here that a critical feature of the fireball-shock model is that the temporal index $\alpha$ is the same across all frequency bands.

We use a Bayesian framework to fit the fireball model to the X-ray afterglow data of GRB140903A and GRB130603B. The general form for $N$ power laws is given by
\begin{equation}
L(t) = 
\begin{cases}
A_{1}t^{\alpha_{1}}, & t \leq t_{1} \\
A_{2}t^{\alpha_{2}}, & t_{1} < t \leq t_{2} \\
..., & ... \\
A_{\rm{N}}t^{\alpha_{\rm{N}}}, & t > t_{\rm{N}}
\end{cases},
\end{equation}
where $A_i, \alpha_i$, and $t_i$ are the amplitude, temporal index and time since burst of the $i^{\rm{th}}$ component-break. We note here that the first power law models the prompt emission. To compare directly with the millisecond-magnetar model, we reparameterize the series of broken power laws in terms of $\Delta t_{i}$, which is the time between successive breaks. We explain this point in more detail in Sec. \ref{sec:magnetar}. Bayesian inference requires us to define priors to allow fitting of the model. For the amplitudes, we only require a prior on the first $A_1$ as the others are determined by demanding the light curve be continuous between any two component breaks.  For $A_1$ we use a log-uniform prior between $10^{-1}$ and $10^{5}$ $L_{50}$, where $L_{50}=10^{50}$ erg s$^{-1}$.  For each of the power-law exponents $\alpha_N$, we set a uniform prior between $-10$ and $0$. For the time between successive temporal breaks we use a log-uniform prior between $10^{-10}$ and $10^{6}$~s except for $\Delta t_{2}$ where we set the minimum of the prior on $\Delta t_2$ to 10~s as the first power-law component models the prompt emission, which for short gamma-ray bursts can last up to this time. We can derive the rest of the parameters using these priors. 

The number of power-law components is itself a free parameter to be fit for. In our analysis, we consider a maximum number of components of 6 as with the inclusion of the prompt emission power law, it is difficult to expect more than 4 temporal breaks \citep[e.g.,][]{Sari1998}. We find the addition of more components does not provide a better fit, a point we discuss further below. We fit our model using the nested sampling package {\sc MultiNest} \citep{Feroz2009}, which allows us to evaluate the evidence for our model given the data. Evidence and the basics of Bayesian inference and model selection are explained in Appendix \ref{Appendix}. We iteratively fit power-law components, evaluating the evidence at each iteration. The number of components that best explains the data is given by that with the highest evidence. 
We find that the evidence is maximised with the four component fireball model for both GRB130603B and GRB140903A. In Fig. \ref{Fig. allcurves} we show our fits with the fireball model in the bottom panel, with the evidences $\Z$ shown in Table \ref{table: BF_F/F}. 
\begin{table}[h!]
\centering
\caption{Evidences $\ln \Z$ for GRB130603B and GRB140903A for different power-law components in the fireball model. The subscript denotes the number of power-law components in the fireball model. The model in bold is the favoured number of power-law components for each gamma-ray burst. The evidences and the errors, the latter being the sampling error are both calculated by {\sc MultiNest}.}
\begin{tabular}{||c c c||} 
 \hline
 & GRB130603B & GRB140903A  \\ [0.5ex] 
 \hline\hline
$\ln{} \Z_{1,F}$ & $431 \pm 0.02$ & $57 \pm 0.02$ \\ 
 \hline
$\ln{} \Z_{2,F}$ & $1019 \pm 0.03$ & $434 \pm 0.03$ \\
 \hline
$\ln{} \Z_{3,F}$ & $1258 \pm 0.03$ & $620 \pm 0.03$ \\
 \hline
$\ln{} \Z_{4,F}$ & $\mathbf{1280 \pm 0.03}$ & $\mathbf{637 \pm 0.03}$ \\
 \hline
 $\ln{} \Z_{5,F}$ & $1275 \pm 0.03$ & $637 \pm 0.03$ \\
 \hline
 $\ln{} \Z_{6,F}$ & $1273 \pm 0.04$ & $634 \pm 0.03$ \\ [1ex]
 \hline
\end{tabular}
\label{table: BF_F/F}
\end{table}
For GRB140903A, our results for the temporal-break times and number of power-law components excluding the prompt emission and the power-law exponents are consistent with \cite{Troja2016} who analysed GRB140903A data across multiple wavelength bands. 
Our maximum posterior fit parameters for the four-component fireball model of GRB130603B and GRB140903A are shown in Table \ref{table: bestfitparameters}.
\begin{table}[h!]
\centering
\caption{Maximum posterior parameters of the four-component fireball model for GRB130603B and GRB140903A without the prompt emission power law.}
 \begin{tabular}{||c c c c c c||} 
 \hline
   & $\alpha_{2}$ & $\alpha_{3}$ & $\alpha_{4}$ & $t_{2}$ (s) & $t_{3} $ (s) \\ [0.5ex] 
 \hline\hline
% GRB130603B & $-0.33 \pm 0.08 $ & $-1.28 \pm 0.25$ & $-2.07 \pm 0.13$ & $1459 \pm 544$ & $12180 \pm 433$ \\ 
 GRB130603B & $-0.33 $ & $-1.28 $ & $-2.07$ & $1459$ & $12180$ \\ 
 \hline
% GRB140903A & $-0.15 \pm 0.04$ & $-1.02 \pm 0.08$ & $-1.99 \pm 0.1$ & $5907 \pm 642$ & $45480 \pm 8530$ \\ [1ex]
 GRB140903A & $-0.15$ & $-1.02$ & $-1.99$ & $5907$& $45480$ \\ [1ex]
  \hline
\end{tabular}
\label{table: bestfitparameters}
\end{table}
%%%%%%%%%%%%%%%%%%%%%%%%%%%%%%%%%%%%%%%%%%%%%%%%%

\section{Magnetar model}\label{sec:magnetar}
The millisecond magnetar model was first introduced by \cite{Dai1998} and \cite{Zhang2001} as a model for the X-ray afterglow evolution through sustained energy injection from a millisecond magnetar central engine. \cite{Zhang2001} derived a model for luminosity evolution from the spin down of this millisecond magnetar through magnetic dipole radiation producing the X-ray afterglow.
\cite{DallOsso2011} extended this model to provide a full solution for energy injection from a magnetar central engine that spins down through magnetic dipole radiation while also including effects of radiative losses due to shocks in the interstellar medium. They also showed that this model allows for non-zero slopes in the plateau which is helpful in explaining the observations of several gamma-ray bursts. \citeauthor{Rowlinson2010} (\citeyear{Rowlinson2010, Rowlinson2013}) successfully fit the millisecond magnetar model of \cite{Zhang2001} to various short gamma-ray bursts. \cite{DallOsso2011} and \cite{Stratta2018} fit the model from \cite{DallOsso2011} to a sample of long and short gamma-ray bursts. 
\cite{LaskyLeris2017} extended the millisecond magnetar model to include the spin down of magnetars with arbitrary braking indices $n$, and fit this model to the X-ray afterglows of GRB130603B and GRB140903A in a Bayesian framework.
\cite{LaskyLeris2017} measured the braking index for both GRB130603B and GRB140903A as $n = 2.9 \pm 0.1$ and $n = 2.6 \pm 0.1$ respectively; the former being consistent with the $n = 3$ value associated with a star spinning down predominantly through magnetic dipole radiation. The braking index was fixed as $n = 3$ in the fits of \cite{Rowlinson2013} and \cite{Zhang2017}.

The generalised millisecond magnetar model is \citep{LaskyLeris2017},
\begin{equation}\label{eq. genmillisecondmagnetar}
L(t) = A_{1}t^{\alpha_{1}} + L_{0}\left(1 + \frac{t}{\tau}\right)^{\frac{1 + n}{1 - n}},
\end{equation}
where the first term corresponds to the prompt emission, which is the same as the fireball model described in Sec. \ref{sec:FB}, $L_0$ is the initial luminosity at the onset of the plateau phase, $\tau$ is the spin-down timescale, and $n$ is the braking index which parameterizes the dominant mode of radiation causing spin-down of the millisecond magnetar.

We reparameterize the millisecond magnetar model (Eq. \ref{eq. genmillisecondmagnetar}) to allow direct comparison with the fireball model. This reparameterization is approximately similar as the first term is sub-dominant at later times by several orders of magnitude. The reparameterized millisecond-magnetar is 
\begin{equation}\label{eq. reparammsmag}
L(t) = 
\begin{cases}
A_{1}t^{\alpha_{1}}, & t < t_{1} \\
A_{2}\left(1 + \frac{t}{\tau}\right)^{\alpha}, & t > t_{1}.
\end{cases}
\end{equation}
This implies two different $\alpha$ values for $t > t_{1}$. An $\alpha_{2} \approx 0$ when $t < \tau$ and an $\alpha_{3} = (1 + n)/(1 - n)$ for $t > \tau$.
The reparameterization allows us to define priors for the magnetar model which are equivalent to those used in the fireball model. We can then use our previously defined priors on $\Delta t_{2}$ and $\alpha$ and construct the parameters $\tau$ and $n$ via
\begin{equation}
\tau = \Delta t_{1} + \Delta t_{2},
\end{equation}
and 
\begin{equation}
n = \frac{\alpha - 1}{\alpha + 1}.
\end{equation}

With this reparameterization, the three component fireball model and millisecond magnetar model have the same parameters. 
Implicitly, the two models have the parameters, 
\begin{equation}
\{A_{1}, \alpha_{1}, \Delta t_{1}, \alpha_{2}, \Delta t_{2}, \alpha_{3}\}.
\end{equation}
We note that the millisecond magnetar model does not explicitly have three power-law exponents, however mathematically for $t < \tau$ the millisecond magnetar model has an $\alpha_2 ~\approx ~0$ power-law exponent, while $\alpha_{3}$ is the power-law exponent for $t > \tau$.
We fit this reparameterized millisecond magnetar to the X-ray afterglow of GRB130603B and GRB140903A, our resulting fit to both gamma-ray burst light curves are shown in Fig. \ref{Fig. allcurves}. For both gamma-ray bursts, our reparameterized millisecond magnetar model produces similar posteriors for $\tau$ and $n$ as \cite{LaskyLeris2017}.
%%%%%%%%%%%%%%%%%%%%%%%%%%%%%%%%%%%%%%%%%%%%%%%%%%%%%%%%%%%%%%%%%%%%%%%%%%%%%%%%%
\section{Model Selection}\label{sec:mds}
\begin{figure*}[!htbp]
  \begin{tabular}{cc}     
        \includegraphics[width=1.0\textwidth]{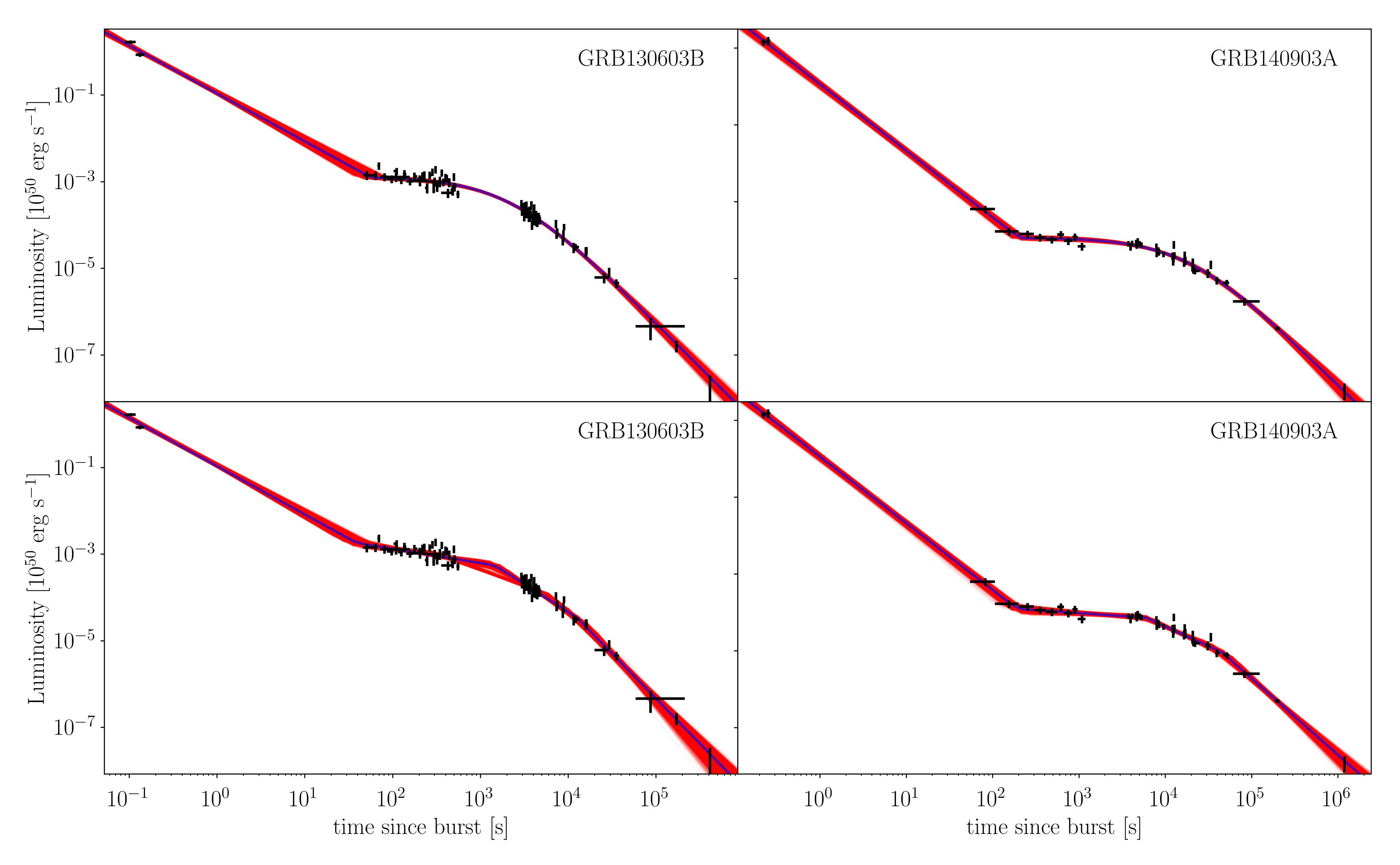} \hspace{-.4cm}
  \end{tabular}
  \caption{X-ray lightcurves for GRB130603B (left panels) and GRB140903A (right panels). Black points indicate data from \textit{Swift} and \textit{Chandra} satellites. The blue curve shows the maximum likelihood model for the millisecond magnetar model (top row) and four-component fireball model (bottom row). The dark red band is the superposition of $5000$ models randomly drawn from the posterior distribution.} 
\label{Fig. allcurves}  
\end{figure*}

In Fig. \ref{Fig. allcurves} we show X-ray lightcurves of GRB130603B (left panels) and GRB140903A (right panels) with the millisecond magnetar model (top row) and the four-component fireball model (bottom row).

We calculate the Bayes factor (see Appendix \ref{Appendix}) to compare between the two models. We find that the Bayes factor $\text{BF}_{\textrm{M/F}} = 19 \ \textrm{and} \ 2271$ for GRB130603B and GRB140903A, respectively. If we assume both hypotheses are equally likely {\it a priori} by setting the prior odds $\Pi_{\textrm{M}}/\Pi_{\textrm{F}}=1$, then this tells us that the data prefers the millisecond magnetar model by $19$ and $2271$ times over a four-component fireball model for GRB130603B and GRB140903A, respectively. Although both models are mathematically similar, the magnetar model is preferred by the data as it provides a smooth transition between power laws compared to the fireball model that has sharp transitions.

For both gamma-ray bursts we find that the millisecond magnetar model is favoured over the fireball model assuming both hypotheses are equally likely {\it a priori}. 
%%%%%%%%%%%%%
\subsection{Prior Odds} \label{sec:odds}
The odds $\mathcal{O}$ is the actual quantity that should be used for doing model selection (see discussion in Appendix \ref{Appendix}). The odds is the product of the Bayes factor and the prior odds; see Eq. (\ref{eq.app:oddsratio}).  Therefore, our analysis in the previous section where we used the Bayes Factor for model selection implicitly set the prior odds as $\Pi_{\textrm{M}}/\Pi_{\textrm{F}}=1$.  In this section, we improve on this by creating an {\it informed} prior on the probability that a long-lived magnetar exists following the gamma-ray burst.

If we assume all short gamma-ray burst progenitors are binary neutron star mergers, then the existence of a long-lived magnetar remnant is dependent on the masses of the progenitor and the nuclear equation of state.
The equation of state dictates the maximum possible non-rotating mass of a neutron star, otherwise known as the Tolman-Oppenheimer-Volkoff mass $M_{\textrm{TOV}}$ \citep{Tolman1939, Oppenheimer1939}. 
The most massive neutron star observed to date has a mass of $2.01 M_\odot$~\citep{Antoniadis2013}, which is therefore the smallest possible value of $M_{\textrm{TOV}}$ given current observations of neutron stars. 
For a millisecond magnetar to be stable and not collapse to a black hole, its mass must be less than $M_{\textrm{TOV}}$. However, millisecond magnetars are born rapidly rotating and can often be supramassive neutron stars which have masses up to $\sim 1.2 \ \times$ $M_{\textrm{TOV}}$ \citep{Cook1994}.
However, supramassive neutron stars collapse on timescales between $\sim 10 - 10^{4}$ s \citep{Ravi2014}. As both GRB130603B and GRB140903A have observations lasting longer than $\sim 10^{4}$ s, if they are millisecond magnetars, they must be infinitely stable neutron stars with mass less than $M_{\textrm{TOV}}$.

We estimate the fraction of binary neutron star mergers that result in an infinitely stable neutron star remnant as follows.
Following~\cite{Lasky2014}, we calculate the post-merger mass distribution $P(\textrm{M})$ based off the statistically-determined mass distribution of galactic binary neutron star systems calculated by~\cite{Kiziltan2013} $M = 1.32 \pm 0.11 M_{\odot}$. 
A calculation of the post-merger mass distribution then requires conversion of the gravitational mass to the rest mass. An approximate relation to make this conversion is, $M_{\text{rest}} = M + 0.075M^{2}$ \citep{Timmes1996}.~Conservation of rest mass in the merger then leads to the post-merger mass distribution.
In reality, the calculation needs to account for the mass ejected during merger. 
Numerical simulations of binary neutron star mergers indicate that the mass ejected is $\lesssim 0.01 M_{\odot}$ \citep[e.g.,][]{Hotokezaka2013, Giacomazzo2013}. 
However, this is inconsistent with observations of the electromagnetic transient to GW170817, which requires a mass ejecta $\sim 0.03 M_{\odot}$ to explain the blue kilonova \citep[e.g.,][]{Evans2017}, while an ejecta mass $\approx 0.07 M_{\odot}$ is required to explain both the blue and red kilonova observations \citep[e.g.,][]{Metzger2017}.
Accounting for the mass ejecta as indicated by the blue and red kilonova observations of GW170817 leads to the post-merger mass distribution of $P(M)= 2.38 \pm 0.14 M_{\odot}$, while ignoring the ejected mass gives $P(M)= 2.45 \pm 0.14 M_{\odot}$. We note that the post-merger mass distribution calculated in \cite{Lasky2014} is incorrectly written as an asymmetrical distribution due to a rounding error.

The prior odds for magnetar vs. fireball models is then the probability that the post-merger mass is less than the (unknown) maximum non-rotating mass $M_{\rm TOV}$.  That is, the prior odds is 
\begin{equation}\label{eq. M/F:Odds}
\frac{\Pi_{{\textrm{M}}}}{\Pi_{\textrm{F}}} = {\int^{M_{\textrm{TOV}}}_{0}P(M) \ dM}.
\end{equation}

\begin{figure}[!htbp]
\centering
\includegraphics[width=0.5\textwidth]{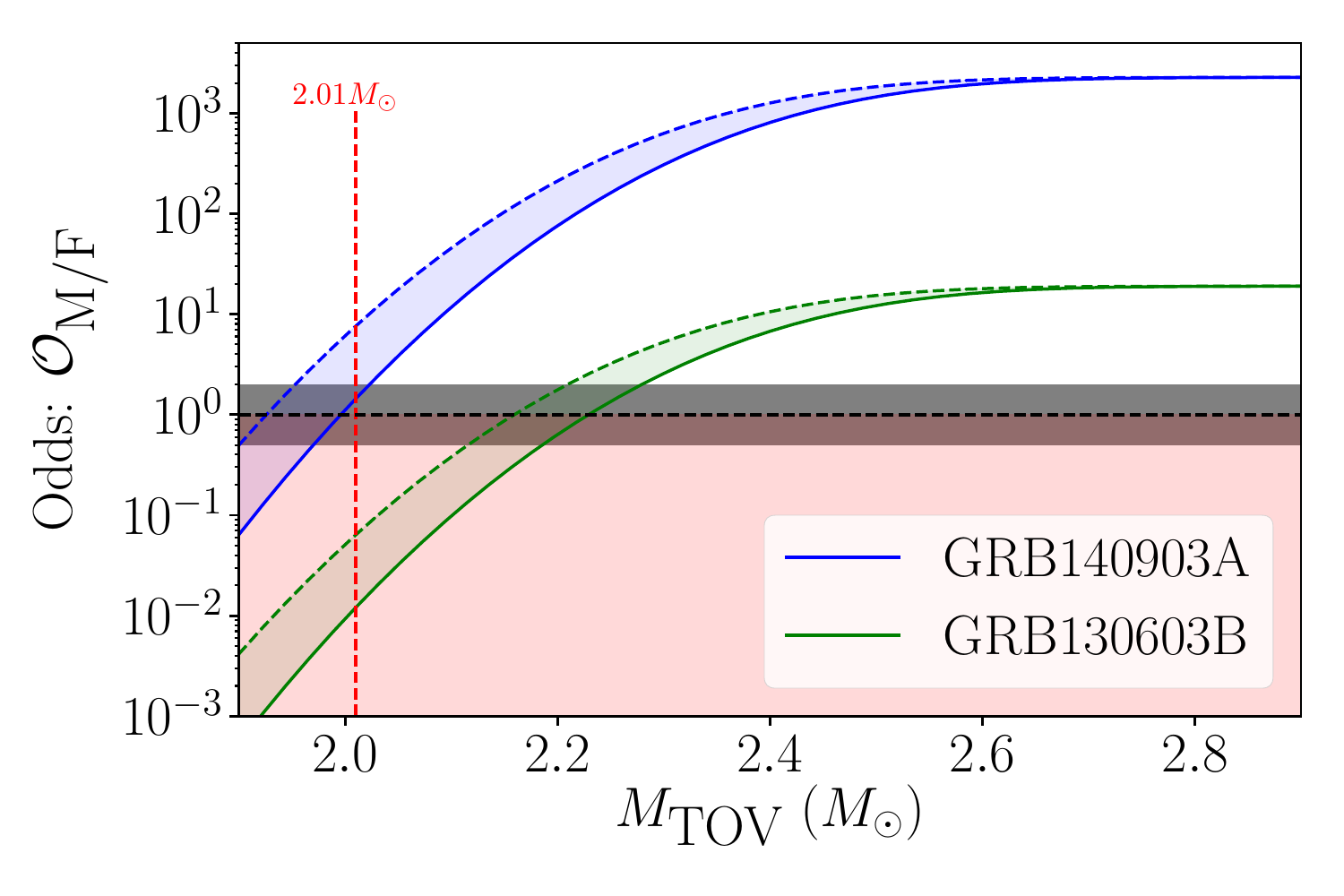}
\caption{Odds $\mathcal{O}_{{\textrm{M/F}}}$ as a function of $M_{\textrm{TOV}}$. The blue and green curves correspond to GRB140903A and GRB130603B, respectively, with the solid curves ignoring mass ejected in mergers, while dashed curves correspond to $0.07 M_{\odot}$ consistent with the blue and red kilonova observations of GW170817 \citep[e.g.,][]{Evans2017, Metzger2017}. The blue and green shaded regions correspond to odds for ejecta masses between $0 M_{\odot}$ and $0.07 M_{\odot}$. The red shaded region indicates odds where the fireball model is favoured over the millisecond magnetar model. The dotted red curve corresponds to a $2.01 M_{\odot}$ which is the smallest possible value of $M_{\textrm{TOV}}$ given current observations of neutron stars. The black dotted line indicates $\mathcal{O} = 1$ while the black shaded region spans an odds of $0.5$ to $2$.}
\label{fig. oddsfunction}
\end{figure}

We use Eq. (\ref{eq. M/F:Odds}) to evaluate the odds $\mathcal{O}_{{\textrm{M/F}}}$ as a function of $M_\textrm{{TOV}}$, which is shown in Fig.~\ref{fig. oddsfunction}.
The odds increases with $M_{\textrm{TOV}}$ as the probability of forming an infinitely stable millisecond magnetar increases for higher $M_{\textrm{TOV}}$. 
The black shaded region indicates an odds confidence interval and spans an odds of $0.5$ to $2.0$.

With an {\it informed} prior odds, selecting between the two models becomes highly dependent on $M_{\textrm{TOV}}$ which is not known. Figure \ref{fig. oddsfunction} shows that for $M_{\textrm{TOV}} \gtrsim 2.01 M_{\odot} \ \textrm{and} \ \gtrsim 2.3 M_{\odot}$ the odds $\mathcal{O}_{{\textrm{M/F}}} \gtrsim 2$, indicating that the millisecond-magnetar model is at least twice as likely than the fireball model for GRB140903A and GRB130603B, respectively.
%%%%%%%%%%%%%
\section{Conclusion} \label{sec:conclusion}
We analyse two short gamma-ray bursts, GRB130603B and GRB140903A and find that the millisecond-magnetar model is favoured over the fireball model with $\text{BF}_{{\textrm{M/F}}} \sim 20$ and $\sim 2270$ respectively, assuming the formation of a millisecond magnetar is just as likely as a fireball. 
When we use an {\it informed} prior odds based on the probability that a long-lived millisecond magnetar is born in the gamma-ray burst, model selection becomes strongly dependent on the maximum allowed non-rotating mass $M_{\textrm{TOV}}$, which is not known. However, we show that GRB140903A favours the millisecond-magnetar model for $M_{\textrm{TOV}} \gtrsim 2.01 \ M_{\odot}$ which is also the most massive neutron star observed to date \citep{Antoniadis2013}. Therefore, for all possible equation of states, GRB140903A favours the millisecond-magnetar model. Similarly, GRB130603B favours the millisecond magnetar model for $M_{\textrm{TOV}} \gtrsim 2.3 \ M_{\odot}$.

Our results show that for GR140903A and GRB130603B the millisecond-magnetar model is favoured over the fireball-shock model solely in the context of the X-ray afterglow data considered here for conservative assumptions on the value of $M_{\textrm{TOV}}$. This has significant implications for gamma-ray burst physics and the fate of the post-merger remnant of neutron star mergers. 
The millisecond-magnetar model implies a central neutron star engine that will spin down and emit gravitational waves, which may be detectable with current and future generation of gravitational-wave detectors \citep[e.g.,][]{Stella2005, DallOsso2007, DallOsso2009, Corsi2009, DallOsso2015, Doneva2015, Lasky2016, Lu2016, Wynn2016, Piro2017, Gao2017, Sarin2018, Lu2018, DallOsso2018}.
Motivated by the possibility of a long-lived neutron star post-merger remnant as supported by the kilonova observations \citep{Yu2018, Ai2018} and an X-ray excess in the afterglow of GRB170817A \citep{Piro2018}, a search for gravitational-wave signals from a possible post-merger remnant from GW170817 was performed, with no detection as expected by theoretical constraints and current detector sensitivities \citep{postmerger2017, postmerger2018}. The putative neutron star born in GRB170817A could also have collapsed to form a black hole instantly, which constrains $M_{\textrm{TOV}} \lesssim 2.16 \ M_{\odot}$ \citep[e.g,][]{Ruiz2018,Rezzolla2018}.

In the future, we aim to extend this analysis to a population of short gamma-ray burst afterglows. Studying an entire population will allow a similar analysis to be performed as in Sec. \ref{sec:mds} allowing us to probe the equation of state through this method.
%%%%%%%%%%%%%%%%%%%%%%%%%%%%%%%%%%%%%%%%%%%%%%%%%%%%%%%%%%%%%%%%%%%%%%%%%%%%%%%%
\appendix
\section{Bayesian Inference}\label{Appendix}
The foundation of this work is based on Bayesian inference. Given a model $\M$ with an associated set of parameters $\theta$, Bayes theorem allows one to calculate the posterior distributions on the model parameters $p(\theta|d, \M)$ given data $d$
\begin{align}
    p(\theta|d,\M)=\frac{\L(d|\theta,\M)\pi(\theta|\M)}{\Z(d|\M)},\label{eq:bayes}
\end{align}
where $\L(d|\theta,\M)$ is the likelihood of the data given the model parameters, $\pi(\theta|\M)$ reflects our prior knowledge of the model parameters, and $\Z(d|\M)$ is the evidence
\begin{align}
    \Z(d|\M)=\int d\theta\L(d|\theta,\M)\pi(\theta|\M).\label{eq:evidence}
\end{align}
The evidence plays no role when estimating the parameters for the model, however, it can be used to do a model comparison between two hypotheses, $\M_1$ and $\M_2$.  In that case, the ratio of the two evidences is known as the Bayes factor
\begin{align}
    {\rm BF_{\M_{1}/\M_{2}}}=\frac{\Z(d|\M_1)}{\Z(d|\M_2)}.
\end{align}
One can use the Bayes factor to distinguish between two different models or hypotheses using an odds.
\begin{equation}\label{eq.app:oddsratio}
\mathcal{O}_{1/2} = \frac{\Z_{1}}{\Z_{2}}\frac{\Pi_{1}}{\Pi_{2}}.
\end{equation}
Here $\Pi_{1}/\Pi_{2}$ is referred to as the prior odds and describes our prior belief about the relative likelihood of one hypothesis over another.
An odds can then be used for model comparison, an $\mathcal{O}_{1/2} > 1$ indicates that the first hypothesis is favoured over the second hypothesis, while $\mathcal{O}_{1/2} < 1$ indicates the second hypothesis is favoured over the first.
%%%%%%%%%%%%%%%%%%%%%%%%%%%%%%%%%%%%%%%%%%%%%%%%%%%%%%%%%%%%%%%%%%%%%%%%%%%%%%%%
\section{Acknowledgments}
We are grateful to Antonia Rowlinson, Eric Howell, Eric Thrane, Kendall Ackley, and the anonymous reviewer for comments on the manuscript. N.S. is supported through an Australian Postgraduate Award. P.D.L. is supported through Australian Research Council Future Fellowship FT160100112 and ARC Discovery Project DP180103155. 
\bibliographystyle{apsrev4-1} 
\bibliography{ref}
\end{document}